\documentclass[a4paper,twoside]{article}
\usepackage{ifthen}
\usepackage{amsmath}
\usepackage[latin1]{inputenc}

\newcommand{\stopbox}{\hspace{\stretch{1}} $\diamond$}
\newcommand{\concept}[1]{\textbf{\textit{#1}}}

\newcommand{\refeqn}[1]{(\ref{#1})}

\newcommand{\T}[2][{}]{T^{\mathcal{A}}{}_{#2}{}^{#1}}
\newcommand{\U}{\mathsf{U}}
\newcommand{\V}{\mathsf{V}}
\newcommand{\W}{\mathsf{W}}

\newcommand{\se}[1]{\mathcal{#1}}

\newcommand{\gd}{\delta}

\newcommand{\chain}[5]{\underset{#1}{#2}[#3]^{#4}{}_{#5}}

\newcommand{\reftitle}[1]{\textit{#1}}
\newcommand{\refarticle}[1]{``#1''}
\newcommand{\refjour}[1]{\textit{#1}}
\newcommand{\volume}[1]{\textbf{#1}}
\newcommand{\isbn}[1]{ISBN #1}

\newtheorem{:definition}{Definition}[section]
\newtheorem{:theorem}[:definition]{Theorem}
\newtheorem{:lemma}[:definition]{Lemma}
\newtheorem{:proof}{Proof}
\newtheorem{:corollary}[:definition]{Corollary}
\newtheorem{:example}[:definition]{Example}

\newenvironment{definition}[1][:Default]
  {\ifthenelse{\equal{#1}{:Default}}
    {\begin{:definition}}
    {\begin{:definition}[#1]}}
  {\stopbox \end{:definition}}

\newenvironment{theorem}[1][:Default]
  {\ifthenelse{\equal{#1}{:Default}}
    {\begin{:theorem}}
    {\begin{:theorem}[#1]}}
  {\stopbox \end{:theorem}}

\newenvironment{lemma}[1][:Default]
  {\ifthenelse{\equal{#1}{:Default}}
    {\begin{:lemma}}
    {\begin{:lemma}[#1]}}
  {\stopbox \end{:lemma}}

\newenvironment{proof}[1][:Default]
  {\ifthenelse{\equal{#1}{:Default}}
    {\paragraph{\textbf{Proof}}}
    {\paragraph{\textbf{Proof (#1)}}}}
  {\stopbox \vspace{1em} \linebreak }

\newenvironment{corollary}[1][:Default]
  {\ifthenelse{\equal{#1}{:Default}}
    {\begin{:corollary}}
    {\begin{:corollary}[#1]}}
  {\stopbox \end{:corollary}}

\begin{document}
%
%
%
%
%
%

\title{Highly structured tensor identities for (2,2)-forms in four
  dimensions.} \author{Ola Wingbrant\footnote{Valhallagatan 26B SE-582
    43 Linköping\newline E-mail: ola@wingbrant.info}} \maketitle

\begin{abstract}
In an $n$ dimensional vector space, any tensor which is antisymmetric
in $k>n$ arguments must vanish; this is a trivial consequence of the
limited number of dimensions. However, when other possible properties
of tensors, for example trace-freeness, are taken into account, such
identities may be heavily disguised. Tensor identities of this kind
were first considered by Lovelock, and later by Edgar and Höglund.
  
In this paper we continue their work. We obtain dimensionally
dependent identities for highly structured expressions of products of
(2,2)-forms. For tensors possessing more symmetries, such as block
symmetry $W_{abcd} = W_{cdab}$, or the first Bianchi identity
$W_{a[bcd]} = 0$, we derive identities for less
structured expressions.
  
These identities are important tools when studying super-energy
tensors, and, in turn, deriving identities for them. As an application
we are able to show that the Bel-Robinson tensor, the super-energy
tensor for the Weyl tensor, satisfies the equation
$\se{T}_{abcy}\se{T}^{abcx} =
\frac{1}{4}\gd^{x}_{y}\se{T}_{abcd}\se{T}^{abcd}$ in four dimensions,
irrespective of the signature of the space.
\end{abstract}


\section{Introduction}
The natural language for general relativity in four dimensions, and
with a space having Lorentzian signature, is spinors
\cite{Penrose-Rindler-1}. The dimension and signature is built into
the spinor formalism, and this is both its strength and its weakness.
Many important relations are obvious or easy to derive in spinors,
while their tensor counterparts are more difficult.

However, if we want to go beyond four dimensions and/or use metrics
with signatures other than Lorentzian, the tensor formalism is the
usual choice at hand. So in these spaces other means to obtain such
important relations have to be used. One method for deriving tensor
relations that has proven useful is to exploit so called
``dimensionally dependent identities''
\cite{Bergqvist02,Edgar02,Lovelock67,Lovelock70}.

This method is built around the process of antisymmetrizing over more
arguments than the number of dimensions of the space. It was first
considered by David Lovelock in the late 1960's
\cite{Lovelock67,Lovelock70}, when he drew attention to a set of
apparently unrelated tensor identities used in general relativity. He
managed to show that they were all special cases of two basic
identities, both having a very simple structure. Lovelock gave two
theorems, and showed that they were consequences of the dimension
alone.

Lovelock's work has recently resurfaced and his theorems have been
generalized by Edgar and Höglund \cite{Edgar02}. We will continue
their work, and show how their ideas can be used in proving a set of
relations for highly structured tensor expressions.

Some of the four-dimensional results presented here are already known
in the special case of a space with Lorentz signature, and are more
or less trivial in the spinor formalism. However, by considering such
relations in tensors instead, we are guided in the derivation of the
corresponding tensor relations. This knowledge may then be used when
trying to prove, or refute, the existence of analogous tensor
relations in higher dimensions \cite{Edgar03}.

A wide class of tensors that have been shown to be significant in both
four and higher dimensions are the super-energy tensors; see
\cite{Senovilla} and references therein. One example is the
Bel-Robinson tensor \cite{Bel58}, which was originally constructed in four
dimensions as the super-energy tensor for the gravitational field.
However, it has subsequently been used in, e.g., eleven dimensional
supergravity theories \cite{Deser}.

In addition, Bergqvist \cite{Bergqvist99} has shown that, in four
dimensions, certain properties of super-energy tensors are
investigated more easily using the equivalent super-energy spinors.

As an application of the identities we derive, we show how
to obtain a quadratic identity for the Bel-Robinson tensor,
\begin{equation}
\se{T}_{abcy}\se{T}^{abcx} =
\frac{1}{4}\gd^{x}_{y}\se{T}_{abcd}\se{T}^{abcd}
\end{equation}
by purely tensorial means.

In this paper we will make use of the abstract index notation as
defined in \cite{Penrose-Rindler-1}.  We also introduce the convention that
$T$ is any tensor, without any specified trace or symmetry properties.
$U$ will always symbolize a \concept{trace-free (2,2)-form}. That is,
\begin{equation}
  U^{a}{}_{bad} = 0 \qquad
  U_{abcd} = U_{[ab]cd} = U_{ab[cd]}.
\end{equation}
$V$ will be taken to be a \concept{block symmetric, trace-free
  (2,2)-form}, hence
\begin{equation}
  V^{a}{}_{bad} = 0 \qquad
  V_{abcd} = V_{[ab]cd} = V_{ab[cd]} \qquad
  V_{abcd} = V_{cdab}.
\end{equation}
Finally, $W$ will stand for a \concept{Weyl candidate}. A Weyl
candidate is a four index tensor with all the algebraic properties of
the Weyl tensor, and thus
\begin{equation}\label{d-weylCandidate}
  W^{a}{}_{bad} = 0 \quad
  W_{abcd} = W_{[ab]cd} = W_{ab[cd]} \quad
  W_{abcd} = W_{cdab} \quad
  W_{a[bcd]} = 0.
\end{equation}
The last relation of \refeqn{d-weylCandidate} is often referred to as
the \concept{first Bianchi identity}. Note that nothing is
said about the differential properties of $W$. Note also that each
class is a special case of the one before, and hence, results obtained
for $U$ will apply to both $V$ and $W$ etc.

Their counterparts without the trace-free property will be represented
by $\U$, $\V$ and $\W$ respectively.

\section{Basic quadratic identities}\label{s-basicIdentities}
In this section we will follow \cite{Penrose-Rindler-1}
and adopt the convention that $\T{a_1}$ denotes a tensor which, in
addition to the free index $a_1$, has an arbitrary free index configuration
of upper and/or lower indices $\mathcal{A}$.

In an $n$-dimensional vector space, we have
\begin{equation}\label{q-singleIndexAnti}
  \T{[a_1}\gd^{b_2}_{a_2}\gd^{b_3}_{a_3}\ldots\gd^{b_{n+1}}_{a_{n+1}]} = 0,
\end{equation}
since the left hand side is antisymmetric in more indices than the
dimension of the space. By taking more indices explicitly into the
antisymmetrization we are able to obtain other identities, which
contain fewer deltas,
\begin{equation}\label{q-multiIndexAnti-1}
  \T{[a_1\ldots a_k}\gd^{b_{k+1}}_{a_{k+1}}
  \gd^{b_{k+2}}_{a_{k+2}}\ldots\gd^{b_{n+1}}_{a_{n+1}]} = 0.
\end{equation}
Such identities would only involve the antisymmetric part of $T$ with
respect to the explicit indices. However, when $T$ itself is antisymmetric in
these indices, equation \refeqn{q-multiIndexAnti-1} will give us an
identity with fewer deltas than \refeqn{q-singleIndexAnti}, but without
losing any part of $T$.

The arguments work equally well for $(k,l)$-forms. Hence,
\begin{equation}\label{q-multiIndexAnti-2}
  \T[{[b_1\ldots b_l}]{a_1\ldots a_k}\gd^{b_{k+1}}_{a_{k+1}}
  \gd^{b_{k+2}}_{a_{k+2}}\ldots\gd^{b_{n+1]}}_{a_{n+1}} = 0
\end{equation}
when $l \geq k$.

These ideas have been thoroughly investigated, and the next theorem is
due to \cite{Edgar02}.

\begin{theorem}\label{t-masterTheorem}
  In an n-dimensional space let $\T[b_1\ldots b_l]{a_1\ldots
    a_k} = \T[{[b_1\ldots b_l]}]{[a_1\ldots a_k]}$ be trace-free on
  its explicit indices. Then
  \begin{equation}
    \T[{[b_1\ldots b_l}]{[a_1\ldots a_k}
    \gd^{b_{l+1}}_{a_{k+1}}\ldots\gd^{b_{l+d}]}_{a_{k+d}]} = 0
  \end{equation}
  where $d+k+l\geq n+1$ and $d \geq 0$.
\end{theorem}
\begin{proof}
  Antisymmetrizing over $k+l+d \geq n+1$ indices, we have
  \begin{eqnarray}
    0 &=& \T[{i_1\ldots i_l}]{[a_1\ldots a_k}
    \gd^{b_{l+1}}_{a_{k+1}}\ldots\gd^{b_{l+d}}_{a_{k+d}}
    \gd^{b_1}_{i_1}\ldots\gd^{b_l}_{i_l]}\\
    &=& \T[{i_1\ldots i_l}]{[a_1\ldots a_k}
    \gd^{[b_{l+1}}_{a_{k+1}}\ldots\gd^{b_{l+d}}_{a_{k+d}}
    \gd^{b_1}_{i_1}\ldots\gd^{b_l]}_{i_l]}
  \end{eqnarray}
  Since $T$ is trace-free on explicit indices, this reduces to
  \begin{equation}
    0 = \T[{i_1\ldots i_l}]{[a_1\ldots a_k}
    \gd^{[b_{l+1}}_{a_{k+1}}\ldots\gd^{b_{l+d}}_{a_{k+d}]}
    \gd^{b_1}_{i_1}\ldots\gd^{b_l]}_{i_l}
  \end{equation}
  The theorem follows by absorbing the deltas.
\end{proof}
Since we will restrict our discussions to $(2,2)$-forms in four
dimensions, Theorem \ref{t-masterTheorem} will not be exploited to its
full extent. In this special case, we have
\begin{corollary}\label{t-mt4D}
  Let $U^{ab}{}_{cd}$ be a trace-free (2,2)-form. In four dimensions,
  \begin{equation}
    U^{[ab}{}_{[cd}\gd^{x]}_{y]} = 0.
  \end{equation}
\end{corollary}

We will now derive three identities, which we will use frequently
throughout the rest of this paper. They are all simple but elegant
consequences of Corollary \ref{t-mt4D}. The most compact presentation
of each of these identities is for trace-free forms; but as will be
seen, alternative presentations are possible without the trace-free
condition.

Our first result is one of the identities considered by Lovelock
\cite{Lovelock70}. We give the proof in detail since it illustrates the
basic idea.

\begin{theorem}\label{t-basic4DId-1}
  Let $U^{ab}{}_{cd}$ be a trace-free (2,2)-form. In four
  dimensions
  \begin{equation}\label{q-basic4DId-1}
    U^{xb}{}_{cd}U^{cd}{}_{yb} = 
    \frac{1}{4}\gd_{y}^{x}U^{ab}{}_{cd}U^{cd}{}_{ab}
  \end{equation}
\end{theorem}

\begin{proof}
  By Corollary \ref{t-mt4D} we have
  \begin{equation}\label{q-basic4DId-0}
    U{}^{[ab}{}_{[cd}\gd_{y]}^{x]} = 0.
  \end{equation}  
  Expanding the left hand side gives 36 terms. Using the antisymmetries
  of $U$, this is reduced by a factor 4 to 9 terms:
  \begin{eqnarray}
    0 & = & 
    U^{xb}{}_{yd}\gd^{a}_{c}+U^{xb}{}_{cy}\gd^{a}_{d}+
    U^{ax}{}_{yd}\gd^{b}_{c}+U^{ax}{}_{cy}\gd^{b}_{d}-\nonumber\\
    &&-U^{xb}{}_{cd}\gd^{a}_{y}-U^{ax}{}_{cd}\gd^{b}_{y}-
    U^{ab}{}_{yd}\gd^{x}_{c}-U^{ab}{}_{cy}\gd^{x}_{d}+\nonumber\\
    &&+U^{ab}{}_{cd}\gd^{x}_{y}.
  \end{eqnarray}
  Multiplication by $U^{cd}{}_{ab}$ and using the trace-free property yields
  \begin{equation}
    -4U^{xb}{}_{cd}U^{cd}{}_{yb}
    +\gd_{y}^{x}U^{ab}{}_{cd}U^{cd}{}_{ab} = 0.
  \end{equation}
\end{proof}
Note that if equation \refeqn{q-basic4DId-0} is multiplied by
$U_{ab}{}^{cd}$ instead, we get 
\begin{equation}\label{q-basic4DId-1b}
  U^{xb}{}_{cd}U_{yb}{}^{cd} = 
  \frac{1}{4}\gd_{y}^{x}U^{ab}{}_{cd}U_{ab}{}^{cd}.
\end{equation}
If $U_{abcd}$ is also block symmetric, then \refeqn{q-basic4DId-1b} will of
course be equal to \refeqn{q-basic4DId-1}.

Now, suppose that $U_{abcd}$ is the trace-free part of some (2,2)-form
$\U_{abcd}$, and defined by the equation
\begin{eqnarray}\label{q-Vdef-2}
  U_{abcd} = \U_{abcd}
  -\frac{2}{(n-2)}\left(g_{a[c}\U_{d]b}-g_{b[c}\U_{d]a}\right)
  +\frac{2}{(n-2)(n-1)}\U g_{a[c}g_{d]b}
\end{eqnarray}
where $\U_{ab} = \U^{k}{}_{akb}$ and $\U = \U^{k}{}_{k}$. We may
then substitute $U_{abcd}$ for \refeqn{q-Vdef-2} in
\refeqn{q-basic4DId-1}, and we have
\begin{corollary}\label{q-basic4DId-4b}
  In four dimensions
  \begin{eqnarray}\label{q-basic4DId-4}
    &&\U^{xb}{}_{cd}\U^{cd}{}_{yb}
    +\U \, \U^{x}{}_{y}
    -2\U^{x}{}_{c}\U^{c}{}_{y}
    -2\U^{xb}{}_{yd}\U^{d}{}_{b}=\\
    &&\qquad \frac{1}{4}\gd^{x}_{y}\
    \left(\U^{ab}{}_{cd}\U^{cd}{}_{ab}+\U \, \U
    -4\U^{b}{}_{d}\U^{d}{}_{b}\right)\nonumber
  \end{eqnarray}
  for any (2,2)-form $\U^{ab}{}_{cd}.$
\end{corollary}
If in addition, the tensor is block symmetric, we are able to derive 
\begin{theorem}\label{t-basic4DId-2}
  Let $V_{abcd}$ be a trace-free, block symmetric (2,2)-form.
  In four dimensions
  \begin{equation}\label{q-basic4DId-2}
    V^{x}{}_{bcd}V^{bcd}{}_{y} = 
    \frac{1}{4}\gd_{y}^{x}V^{a}{}_{bcd}V^{bcd}{}_{a}
  \end{equation}
\end{theorem}
The proof is completely analogous to that of Theorem
\ref{t-basic4DId-1} and therefore omitted.
The result may also be found with slightly more work by considering
the identity
\begin{equation}
  V^{[a}{}_{[bcd}\gd^{x]}_{y]}V_{a}{}^{bcd} = 0, 
\end{equation}
which holds by Theorem \ref{t-masterTheorem}.

In the same way as for Theorem \ref{t-basic4DId-1} we may think of
$V_{abcd}$ as the trace-free part of a block symmetric (2,2)-form
$\V_{abcd}$. We have
\begin{corollary}
  For any block symmetric (2,2)-form $\V_{abcd}$,
  \begin{eqnarray}\label{q-basic4DId-3}
    &&\V_{abc}{}^{x}\V^{acb}{}_{y}
    -\V^{x}{}_{a}\V^{a}{}_{y}
    -\V^{xc}{}_{yb}\V^{b}{}_{c}
    +\frac{1}{2}\V^{x}{}_{y}\V=\\
    &&\quad \frac{1}{4}\gd^{x}_{y}\left(
      \V_{abc}{}^{d}\V^{acb}{}_{d}
    -2\V^{c}{}_{b}\V^{b}{}_{c}
    +\frac{1}{2}\V \, \V\right)\nonumber
  \end{eqnarray}
\end{corollary}
This relation will prove to be useful in Section \ref{s-quarticId}.

The next lemma is a bit different; it is concerned with an expression
with four free indices. 
\begin{lemma}[Index switch]\label{t-switchId-1}
  Let $U_{abcd}$ be a trace-free (2,2)-form.
  In four dimensions
  \begin{equation}\label{q-switchId-1}
    2U^{b[w}{}_{cy}U^{x]c}{}_{zb} =
    \frac{1}{2}U^{wx}{}_{bc}U^{bc}{}_{yz}
    -\frac{1}{4}\gd_{y}^{[w}\gd_{z}^{x]}U^{ab}{}_{cd}U^{cd}{}_{ab}
  \end{equation}
\end{lemma}

\begin{proof}
  Consider the identity
  \begin{equation}
    U^{[wb}{}_{[cd}\gd^{x]}_{y]}U^{cd}{}_{zb} = 0.
  \end{equation}
  Expansion and use of equation \refeqn{q-basic4DId-1} gives
  \begin{eqnarray}
    0 & = &
    -\frac{1}{4}\gd_{y}^{x}\gd_{z}^{w}U^{ab}{}_{cd}U^{cd}{}_{ab}
    +\frac{1}{4}\gd_{y}^{w}\gd_{z}^{x}U^{ab}{}_{cd}U^{cd}{}_{ab}-\nonumber\\
    &&-U^{wx}{}_{cd}U^{cd}{}_{yz}
    +2U^{bw}{}_{cy}U^{xc}{}_{zb}
    -2U^{bx}{}_{cy}U^{wc}{}_{zb},
  \end{eqnarray}
  which is \refeqn{q-switchId-1}.
\end{proof}
Note that relation \refeqn{q-switchId-1} may be put in the form
\begin{equation}
  U^{bw}{}_{cy}U^{xc}{}_{zb} = U^{bx}{}_{cy}U^{wc}{}_{zb}
  +\frac{1}{2}U^{wx}{}_{cd}U^{cd}{}_{yz}
  -\frac{1}{4}\gd_{y}^{[w}\gd_{z}^{x]}U^{ab}{}_{cd}U^{cd}{}_{ab}.
\end{equation}
Hence it may be interpreted as a switch of the abstract indices $w$
and $x$.

This lemma will be an invaluable tool in proving Lemma
\ref{t-brokenChain2Id-1}, which in turn, is essential in showing
one of our main results, Theorem \ref{t-quadraticBR-1}.

\section{Chains of trace-free (2,2)-forms}\label{s-chains}
There are some very highly structured product constructions for
$(2,2)$-forms. We will call them \concept{chains} because of the way the
tensors are linked to each other. We will call the number of
participating tensors in a chain the chains \concept{length}.

\begin{definition}[Chain of the zeroth kind]
  An expression of the form
  \begin{equation}
    \underbrace{T^{wx}{}_{c_1d_1}T^{c_1d_1}{}_{c_2d_2}T^{c_2d_2}{}_{c_3d_3}
      \cdots
      T^{c_{m-2}d_{m-2}}{}_{c_{m-1}d_{m-1}}T^{c_{m-1}d_{m-1}}{}_{yz}}_{m}
  \end{equation}
  where indices $w,x,y$ and $z$ are free, is said to be a \concept{chain of
    the zeroth kind} of length $m$ and is written $\chain{0}{T}{m}{wx}{yz}$.
\end{definition}

\begin{theorem}\label{t-chain1Id-1}
  Let $U^{ab}{}_{cd}$ be a trace-free (2,2)-form. In four dimensions
  \begin{equation}\label{q-chain1Id-1}
    \chain{0}{U}{m}{xb}{yb} = \frac{1}{4}\gd^{x}_{y}\chain{0}{U}{m}{ab}{ab}
  \end{equation}
\end{theorem}

\begin{proof}
  The proof is by induction over the number of factors. We know that
  the case $m = 2$ is true by equation \refeqn{q-basic4DId-1}. Assume
  that \refeqn{q-chain1Id-1} holds for $m = k-1 \geq 2$. Consider the
  identity
  \begin{equation}\label{q-chain1Id-3}
    U{}^{[ab}{}_{[cd}\gd_{y]}^{x]}\chain{0}{U}{k-1}{cd}{ab} = 0.
  \end{equation}
  Expansion gives,
  \begin{eqnarray}
    0 & = & U^{ab}{}_{cd}\gd^{x}_{y}\chain{0}{U}{k-1}{cd}{ab}+\\
    &&+(U^{xb}{}_{yd}\gd^{a}_{c}+U^{xb}{}_{cy}\gd^{a}_{d}+
    U^{ax}{}_{yd}\gd^{b}_{c}+U^{ax}{}_{cy}\gd^{b}_{d})
    \chain{0}{U}{k-1}{cd}{ab}+\nonumber\\
    &&+(-U^{xb}{}_{cd}\gd^{a}_{y}-U^{ax}{}_{cd}\gd^{b}_{y}-
    U^{ab}{}_{yd}\gd^{x}_{c}-U^{ab}{}_{cy}\gd^{x}_{d})
    \chain{0}{U}{k-1}{cd}{ab}\nonumber
  \end{eqnarray}
  After renaming dummy indices we are left with 
  \begin{eqnarray}\label{q-chain1Id-2}
    \gd_{y}^{x}U^{ab}{}_{cd}\chain{0}{U}{k-1}{cd}{ab} = 
    4U^{xb}{}_{cd}\chain{0}{U}{k-1}{cd}{yb}
    -4U^{xb}{}_{yd}\chain{0}{U}{k-1}{cd}{cb}
  \end{eqnarray}
  However, the last term of equation \refeqn{q-chain1Id-2} vanishes
  since,
  \begin{eqnarray}\label{q-chain1Id-2b}
    U^{xb}{}_{yd}\chain{0}{U}{k-1}{cd}{cb} = 
    U^{xb}{}_{yd}\gd^{d}_{b}\chain{0}{U}{k-1}{ca}{ca} = 0
  \end{eqnarray}
  The first equality of \refeqn{q-chain1Id-2b} holds by the induction
  hypothesis, and the second one follows from the trace-free property.
  Thus we are left with
  \begin{eqnarray}
    \gd_{y}^{x}U{}^{ab}{}_{cd}\chain{0}{U}{k-1}{cd}{ab}
    & = & 4U^{xb}{}_{cd}\chain{0}{U}{k-1}{cd}{yb}
  \end{eqnarray}
  which is \refeqn{q-chain1Id-1}.
\end{proof}

\begin{definition}[Chain of the first kind]
  An expression of the form
  \begin{equation}
    \underbrace{T^{wc_1}{}_{yd_1}T^{d_1c_2}{}_{c_1d_2}T^{d_2c_3}{}_{c_2d_3}
      \cdots
      T^{d_{m-2}c_{m-1}}{}_{c_{m-2}d_{m-1}}T^{d_{m-1}x}{}_{c_{m-1}z}}_{m}
  \end{equation}
  where indices $w,x,y$ and $z$ are free, is said to be a \concept{chain of
    the first kind} of length $m$, and is written $\chain{1}{T}{m}{wx}{yz}$.
\end{definition}
Note that when $T$ is a $(2,2)$-form $\U$, we see directly from the
definition that
\begin{equation}\label{q-chain2Property-1}
  \chain{1}{\U}{m}{wx}{yz} = \chain{1}{\U}{m}{xw}{zy}
\end{equation}

\begin{theorem}\label{t-chain2Id-1}
  Let $U^{ab}{}_{cd}$ be a trace-free (2,2)-form. In four
  dimensions
  \begin{equation}\label{q-chain2Id-1}
    \chain{1}{U}{m}{xb}{by} = \frac{1}{4}\gd^{x}_{y}\chain{1}{U}{m}{ab}{ba}
  \end{equation}
\end{theorem}
\begin{proof}
  The proof is similar to the proof of Theorem \ref{t-chain1Id-1}
  with only slight modifications. From \refeqn{q-basic4DId-1} we know
  that \refeqn{q-chain2Id-1} holds in the case $m = 2$. Assume that
  it holds for $m = k-1 \geq 2$, and consider the identity
  \begin{equation}\label{q-chain2Id-2}
    U{}^{[ab}{}_{[cd}\gd_{y]}^{x]}\chain{1}{U}{m-1}{dc}{ba} = 0.
  \end{equation}
  Expanding this relation gives
  \begin{eqnarray}
    0 & = & U^{ab}{}_{cd}\gd^{x}_{y}\chain{1}{U}{k-1}{dc}{ba}+\\
    &&(U^{xb}{}_{yd}\gd^{a}_{c}+U^{xb}{}_{cy}\gd^{a}_{d}+
    U^{ax}{}_{yd}\gd^{b}_{c}+U^{ax}{}_{cy}\gd^{b}_{d})
    \chain{1}{U}{k-1}{dc}{ba}+\nonumber\\
    &&+(-U^{xb}{}_{cd}\gd^{a}_{y}-U^{ax}{}_{cd}\gd^{b}_{y}-
    U^{ab}{}_{yd}\gd^{x}_{c}-U^{ab}{}_{cy}\gd^{x}_{d})
    \chain{1}{U}{k-1}{dc}{ba}\nonumber
  \end{eqnarray}
  We see directly that two of the terms on the second row disappear
  because of the trace-free property. If we rewrite the rest of the
  terms, using property \refeqn{q-chain2Property-1} and renaming dummy
  indices, we have
  \begin{eqnarray}\label{q-chain2Id-3}
    \gd^{x}_{y}U^{ab}{}_{cd}\chain{1}{U}{k-1}{dc}{ba} = 
    4U^{xb}{}_{cd}\chain{1}{U}{k-1}{dc}{by}
    -2U^{xb}{}_{cy}\chain{1}{U}{k-1}{cd}{db}
  \end{eqnarray}
  The last term of equation \refeqn{q-chain2Id-3} vanishes
  since,
  \begin{eqnarray}\label{q-chain2Id-4}
    U^{xb}{}_{cy}\chain{1}{U}{k-1}{cd}{db} = 
    U^{xb}{}_{cy}\gd^{c}_{b}\chain{1}{U}{k-1}{ad}{da} = 0
  \end{eqnarray}
  The first equality of \refeqn{q-chain2Id-4} holds by the induction
  hypothesis, the second one follows from the trace-free property.
  Therefore we are left with
  \begin{eqnarray}
    \gd^{x}_{y}U^{ab}{}_{cd}\chain{1}{U}{k-1}{dc}{ba} = 
    4U^{xb}{}_{cd}\chain{1}{U}{k-1}{dc}{by}
  \end{eqnarray}
  and the proof is finished.
\end{proof}
Next we are concerned with one kind of broken chain structure. It is
a chain of the zeroth kind, but it includes an element that destroys
the ordinary regularity.

\begin{theorem}\label{t-irregularChain-1}
  In four dimensions, with $k > 0$ and $l > 0$,
  \begin{equation}
    \chain{0}{U}{k}{xb}{cd}U^{de}{}_{bf}\chain{0}{U}{m}{fc}{ey}
    =\frac{1}{4}\gd^{x}_{y}
    \chain{0}{U}{k}{ab}{cd}U^{de}{}_{bf}\chain{0}{U}{m}{fc}{ea}
  \end{equation}
  for all trace-free (2,2)-forms $U$.
\end{theorem}

\begin{proof}
  Expansion of the identity
  \begin{equation}
    \chain{0}{U}{k}{xf}{ab}U^{[ab}{}_{[cd}
    \gd^{e]}_{f]}\chain{0}{U}{m}{cd}{ye} = 0
  \end{equation}
  yields
  \begin{eqnarray}
    0&=&\chain{0}{U}{k+m+1}{xf}{yf}
    -2\chain{0}{U}{k}{xf}{bf}\chain{0}{U}{m+1}{be}{ye}-\nonumber\\
    &&-2\chain{0}{U}{k+1}{xf}{bf}\chain{0}{U}{m}{be}{ye}
    -4\chain{0}{U}{k}{xf}{ab}U^{be}{}_{fd}\chain{0}{U}{m}{da}{ey}
  \end{eqnarray}
  The first three terms are chains of the zeroth kind, and products of
  chains of the zeroth kind. Thus, by Theorem \ref{t-chain1Id-1} we
  have
  \begin{eqnarray}\label{q-irregularChain-1}
    &&\chain{0}{U}{k}{xf}{ab}U^{be}{}_{fd}\chain{0}{U}{m}{da}{ey} =
    \frac{1}{16}\gd^{x}_{y}\chain{0}{U}{k+m+1}{af}{af}-\nonumber\\
    &&\quad -\frac{1}{32}\gd^{x}_{y}\chain{0}{U}{k}{af}{af}
    \chain{0}{U}{m+1}{de}{de}
    -\frac{1}{32}\gd^{x}_{y}\chain{0}{U}{k+1}{af}{af}\chain{0}{U}{m}{de}{de}
  \end{eqnarray}
  Taking the trace of equation \refeqn{q-irregularChain-1} and
  multiplication by $\frac{1}{4}\gd^{x}_{y}$ gives
  \begin{eqnarray}
    &&\frac{1}{4}\gd^{x}_{y}\chain{0}{U}{k}{cf}{ab}U^{be}{}_{fd}
    \chain{0}{U}{m}{da}{ec} =
    \frac{1}{16}\gd^{x}_{y}\chain{0}{U}{k+m+1}{af}{af}-\nonumber\\
    &&\quad -\frac{1}{32}\gd^{x}_{y}\chain{0}{U}{k}{af}{af}\chain{0}{U}{m+1}{de}{de}
    -\frac{1}{32}\gd^{x}_{y}\chain{0}{U}{k+1}{af}{af}\chain{0}{U}{m}{de}{de}
  \end{eqnarray}
  and the theorem follows.
\end{proof}
The proof immediately gives us a relation between this more
complicated expression and chains of the zeroth kind.
\begin{corollary}
  In four dimensions
  \begin{eqnarray}
    &&\chain{0}{U}{k}{xf}{ab}U^{be}{}_{fd}\chain{0}{U}{m}{da}{ey} =
    \frac{1}{16}\gd^{x}_{y}\left(
      \chain{0}{U}{k+m+1}{af}{af}-\right.\nonumber\\
    &&\quad \left.-\frac{1}{2}\chain{0}{U}{k}{af}{af}
      \chain{0}{U}{m+1}{de}{de}
      -\frac{1}{2}\chain{0}{U}{k+1}{af}{af}\chain{0}{U}{m}{de}{de}
    \right)
  \end{eqnarray}
\end{corollary}
In Section \ref{s-quarticId} we will explicitly use the special case of
Theorem \ref{t-irregularChain-1} when $k=1$ and $m=2$. We therefore
give this result as a separate corollary.
\begin{corollary}\label{t-irregularChain4}
  In four dimensions
  \begin{equation}
    U{}^{xb}{}_{cd}U{}^{de}{}_{bf}U{}^{cf}{}_{gh}U{}^{gh}{}_{ye} =
    \frac{1}{4}\gd^{x}_{y}U{}^{ab}{}_{cd}U{}^{de}{}_{bf}U{}^{cf}{}_{gh}U{}^{gh}{}_{ae}
  \end{equation}
\end{corollary}

\section{Chains of trace-free block symmetric (2,2)-forms}\label{s-blockSymChain}
What happens if the structures of the previous section are disturbed
in some other way? Is it still possible to find identities? In order to
investigate these questions further we start by looking at a chain of
the first kind. For simplicity we will carry out our arguments in the
case when the chain is of length five, although the results are valid
for chains of arbitrary length.
\begin{equation}
    T^{wb}{}_{yd}T^{de}{}_{bf}T^{fg}{}_{eh}T^{hi}{}_{gk}T^{kx}{}_{iz}
\end{equation}
Let $g$ and $h$ change places in the third $T$, we then have
\begin{equation}
    T^{wb}{}_{yd}T^{de}{}_{bf}T^{f}{}_{he}{}^{g}T^{hi}{}_{gk}T^{kx}{}_{iz}
\end{equation}
We shall call such a construction a \concept{twist}.

From now on, until the end of this section, we will only discuss
trace-free (2,2)-forms enjoying the block symmetric property
$V_{abcd} = V_{cdab}$. Since this class of tensors is a special case
of those of the Section \ref{s-chains}, any $V_{abcd}$ will, of
course, obey the results of that section. However, we are able to
derive additional results, originating from the block symmetry.

We immediately see that
\begin{equation}
    V^{wb}{}_{yd}V^{de}{}_{bf}V^{f}{}_{he}{}^{g}V^{hi}{}_{gk}V^{kx}{}_{iz}
    =
    V^{wb}{}_{yd}V^{de}{}_{bf}V^{fh}{}_{eg}V^{g}{}_{kh}{}^{i}V^{kx}{}_{iz}.
\end{equation}
Thus, twists may move freely along chains, and we may always assume
that the twist is at the end of the chain. We shall write such a
chain of length $m$ as $\chain{t1}{V}{m}{wx}{yz}$. A chain of length
$m$ that contains $n < m$ twists, is denoted by
$\chain{nt1}{V}{m}{wx}{yz}$.

Assume that we have a chain with two twists. Because of what we
stated above we can further assume that they are on neighboring
tensors.
\begin{equation}
  V^{wb}{}_{yd}V^{de}{}_{bf}V^{f}{}_{he}{}^{g}V^{h}{}_{kg}{}^{i}V^{kx}{}_{iz}
  =
  V^{wb}{}_{yd}V^{de}{}_{bf}V^{fh}{}_{eg}V_{hk}{}^{gi}V^{kx}{}_{iz}
\end{equation}
By the block symmetry it is clear that the twists cancel. Hence,
we have the following lemma.

\begin{lemma}[Twist reduction]
  Let $V_{abcd}$ be a trace-free, block symmetric (2,2)-form. Then 
  \begin{equation}
    \chain{nt1}{V}{m}{wx}{yz} = \chain{t1}{V}{m}{wx}{yz}
  \end{equation}
  if the number of twists, $n$, are odd, and
  \begin{equation}
    \chain{nt1}{V}{m}{wx}{yz} = \chain{1}{V}{m}{wx}{yz}
  \end{equation}
  if the number of twists are even.
\end{lemma}
Hence, the number of twists reduce to either zero or one,
depending on whether it is even or odd to start with.

\begin{theorem}\label{t-chain3Id-1}
  In four dimensions,
  \begin{equation}\label{q-chain3Id-1}
    \chain{t1}{V}{m}{xb}{by} = \frac{1}{4}\gd^{x}_{y}\chain{t1}{V}{m}{ab}{ba}
  \end{equation}
  for any trace-free block symmetric (2,2)-form $V$.
\end{theorem}
\begin{proof}
  The proof is analogous to those of Theorem \ref{t-chain1Id-1} and
  Theorem \ref{t-chain2Id-1}, and we will therefore leave out some of the
  details.
  
  Equation \refeqn{q-chain3Id-1} is true for $m=2$ by Theorem
  \ref{t-basic4DId-2}. Assume it is true for $m=k-1 \geq 2$, and
  consider the identity
  \begin{equation}
    V{}^{[ab}{}_{[cd}\gd_{y]}^{x]}
    \chain{1}{V}{k-2}{di}{bk}V^{k}{}_{ai}{}^{c} = 0    
  \end{equation}
  Expansion and use of symmetries, trace properties, and induction
  hypothesis yields
  \begin{equation}
    0 = \chain{1}{V}{k-1}{ai}{ck}V^{k}{}_{ai}{}^{c}\gd^{x}_{y}
    -4V^{xb}{}_{cd}\chain{1}{V}{k-2}{di}{bk}V^{k}{}_{yi}{}^{c},
  \end{equation}
  which is the desired result.
\end{proof}

\section{Chains of Weyl candidates.}\label{s-weylChains}
As we pointed out at the beginning of Section \ref{s-basicIdentities},
Weyl candidates are special cases of block symmetric (2,2)-forms,
and will therefore obey all the results of both Section \ref{s-chains}
and Section \ref{s-blockSymChain}. But, since Weyl candidates also
possess the first Bianchi identity, $W_{a[bcd]} = 0$, we are able to
derive an additional result.  In order to do so, we turn our attention
to a variant of chains of the zeroth kind.

Let us, in analogy with the previous section, study a chain of the
zeroth kind of length five.
\begin{equation}
  T^{wx}{}_{cd}T^{cd}{}_{ef}T^{ef}{}_{gh}T^{gh}{}_{ik}T^{ik}{}_{yz}
\end{equation}
Now, twist $f$ and $h$ in the middle factor,
\begin{equation}
  T^{wx}{}_{cd}T^{cd}{}_{ef}T^{e}{}_{hg}{}^{f}T^{gh}{}_{ik}T^{ik}{}_{yz}
\end{equation}
We note that this may be rewritten as,
\begin{eqnarray}
  &&T^{wx}{}_{cd}T^{cd}{}_{ef}T^{e}{}_{hg}{}^{f}T^{gh}{}_{ik}T^{ik}{}_{yz}
  =\nonumber\\
  &&\quad-T^{wx}{}_{cd}T^{cd}{}_{ef}T^{e}{}_{hg}{}^{f}T^{hg}{}_{ik}T^{ik}{}_{yz}
  =
  T^{wx}{}_{cd}T^{cd}{}_{ef}T^{e}{}_{g}{}^{f}{}_{h}T^{gh}{}_{ik}T^{ik}{}_{yz}
\end{eqnarray}
Evidently, twists gives rise to elements that break the structure in
this case as well. Chains of the zeroth kind containing such elements
are denoted by $\chain{t0}{T}{m}{wx}{yz}$.

\begin{lemma}\label{t-bianchiComp}
  For any Weyl candidate $W_{abcd}$
  \begin{equation}
    W_{a}{}^{e}{}_{b}{}^{f}W_{ef}{}^{cd} =
    \frac{1}{2}W_{ab}{}^{ef}W_{ef}{}^{cd}
  \end{equation}
\end{lemma}
\begin{proof}
  By the first Bianchi identity we have
  \begin{equation}
    W_{ab}{}^{ef} =
    \left(W_{a}{}^{e}{}_{b}{}^{f}-W_{a}{}^{f}{}_{b}{}^{e}\right) =
    2W_{a}{}^{[e}{}_{b}{}^{f]}.
  \end{equation}
  And hence
  \begin{equation}
    W_{a}{}^{e}{}_{b}{}^{f}W_{ef}{}^{cd} = 
    W_{a}{}^{e}{}_{b}{}^{f}W_{[ef]}{}^{cd} =
    W_{a}{}^{[e}{}_{b}{}^{f]}W_{ef}{}^{cd} =
    \frac{1}{2}W_{ab}{}^{ef}W_{ef}{}^{cd}
  \end{equation}
\end{proof}

\begin{theorem}
  For chains of the zeroth kind of any Weyl candidate $W_{abcd}$ of
  length $m$, with $n<m$ number of twists
  \begin{equation}
    \chain{nt0}{W}{m}{wx}{yz} = \frac{1}{2^n}\chain{0}{W}{m}{wx}{yz}.
  \end{equation}
\end{theorem}
The proof is immediate from Lemma \ref{t-bianchiComp}.

\begin{corollary}\label{t-distChain1Id-1}
  \begin{equation}
    \chain{nt0}{W}{m}{xb}{yb} =
    \frac{1}{4 \cdot 2^n}\,\gd^{x}_{y}\chain{0}{W}{m}{ab}{ab}
  \end{equation}
  for chains of Weyl candidates $W_{abcd}$ with $n<m$ number of twists.
\end{corollary}

\section{Quartic identities}\label{s-quarticId}
In this section we will break the chain structures of Section
\ref{s-chains} even further. However, we will specialize our
discussion to chains of
length four.

Consider an expression of a block symmetric double $2$-form
\begin{equation}\label{q-brokenChain1Exp}
  V^{xc}{}_{ef}V^{ef}{}_{ab}V^{a}{}_{cgh}V^{gh}{}_{y}{}^{b}.
\end{equation}
This would be a chain of the zeroth kind if $b$ and $c$ changed places
in the last two factors. In order to make the next theorem slightly
more general, we will rearrange this expression. Using the
block symmetries, we have
\begin{equation}\label{q-brokenRearrange}
  V^{xc}{}_{ef}V^{ef}{}_{ab}V^{a}{}_{cgh}V^{gh}{}_{y}{}^{b} =
  V^{xc}{}_{ef}V_{ab}{}^{ef}V^{a}{}_{cgh}V_{y}{}^{b}{}^{gh}
\end{equation}

\begin{lemma}\label{t-brokenChain1Id-1}
  In four dimensions
  \begin{equation}
    U^{xc}{}_{ef}U_{ab}{}^{ef}U^{a}{}_{cgh}U_{y}{}^{b}{}^{gh} =
    \frac{1}{4}\gd^{x}_{y}
    U^{dc}{}_{ef}U_{ab}{}^{ef}U^{a}{}_{cgh}U_{d}{}^{b}{}^{gh}
  \end{equation}
  for all trace-free (2,2)-forms $U_{abcd}$.
\end{lemma}

\begin{proof}
  The main idea is to let a product of two trace-free (2,2)-forms
  substitute for $\V$ in equation \refeqn{q-basic4DId-3}. Put
  \begin{equation}\label{q-Pdef}
    \V_{abcd} = U_{ab}{}^{ef}U_{cdef}
  \end{equation}
  Since $U$ is a (2,2)-form so is $\V$. In addition $\V$ has the block
  symmetry as well. For the trace of $\V$ we have by equation
  \refeqn{q-basic4DId-1b}
  \begin{equation}\label{q-prodProperty-1}
    \V_{a}{}^{b}{}_{bd} = U_{a}{}^{bef}U_{bdef} =
    \frac{1}{4}g_{ad}U^{cbef}U_{bcef} =
    \frac{1}{4}g_{ad}\V^{cb}{}_{bc}
  \end{equation}
  We may now consider relation \refeqn{q-basic4DId-3}. However, by
  virtue of \refeqn{q-prodProperty-1} this simplifies to
  \begin{equation}
    0 = -\V_{abc}{}^{d}\V^{acb}{}_{d}\gd^{x}_{y}
    +4\V_{abc}{}^{x}\V^{acb}{}_{y}
  \end{equation}
  or equivalently
  \begin{equation}
    \V^{x}{}_{cab}{}\V^{ac}{}_{y}{}^{b} = 
    \frac{1}{4}\gd^{x}_{y}\V^{d}{}_{cab}{}\V^{ac}{}_{d}{}^{b}
  \end{equation}
  Hence,
  \begin{equation}
    U^{x}{}_{cef}U_{ab}{}^{ef}U^{ac}{}_{gh}U_{y}{}^{b}{}^{gh} =
    \frac{1}{4}\gd^{x}_{y}
    U^{d}{}_{cef}U_{ab}{}^{ef}U^{ac}{}_{gh}U_{d}{}^{b}{}^{gh}
  \end{equation}
  as required.
\end{proof}
For completeness we note that the following substitutions are also
possible in \refeqn{q-basic4DId-3}. They are all block symmetric and
the trace has the property \refeqn{q-prodProperty-1}.
\begin{eqnarray}
  \V^{ab}{}_{cd} &=& V^{i[a}{}_{j[c}V^{b]j}{}_{d]i}\\
  \V^{ab}{}_{cd} &=& V_{j}{}^{[a|i|}{}_{[c}V^{b]j}{}_{d]i}\\
  \V^{ab}{}_{cd} &=& V_{ij}{}^{[a}{}_{[c}V_{d]}{}^{b]ij}\\
  \V^{ab}{}_{cd} &=& U_{i}{}^{[a}{}_{j}{}^{b]}U^{i}{}_{[c}{}^{j}{}_{d]}
\end{eqnarray}
The first two yield identical expansions; they may with some
effort, and together with Lemma \ref{t-switchId-1}, be used to derive
Lemma \ref{t-brokenChain1Id-1}. The last two merely reproduce Lemma
\ref{t-brokenChain1Id-1} in a more indirect way. Hence nothing new can
be found from any of these.

\begin{lemma}\label{t-brokenChain2Id-1}
  Let $W_{abcd}$ be a Weyl candidate. In four dimensions
  \begin{equation}\label{q-4brokenChainId-1}
    W{}^{xb}{}_{cd}W{}^{de}{}_{bf}W{}^{fgc}{}_{h}W{}^{h}{}_{egy} = 
    \frac{1}{4}\gd_{y}^{x}
    W{}^{ab}{}_{cd}W{}^{de}{}_{bf}W{}^{fgc}{}_{h}W{}^{h}{}_{ega}
  \end{equation}
\end{lemma}

\begin{proof}
  We note that expression \refeqn{q-4brokenChainId-1} would be a chain
  of the first kind of length four if $c$ and $e$ in the last two
  factors were interchanged. Start with the left-hand-side,
  \begin{eqnarray}
    &&W^{xb}{}_{cd}W^{de}{}_{bf}
    \underbrace{W^{fgc}{}_{h}}_{Bianchi}W^{h}{}_{egy}=\\
    &&=-W^{xb}{}_{cd}W^{de}{}_{bf}
    \underbrace{W^{fc}{}_{h}{}^{g}W^{h}{}_{egy}}_{lem.\ \ref{t-bianchiComp}}
    -W^{xb}{}_{cd}W^{de}{}_{bf}
    \underbrace{W^{f}{}_{h}{}^{gc}W^{h}{}_{egy}}_{lem.\
      \ref{t-switchId-1},
      \,c\,\leftrightarrow\,e}=\\
    &&=-\frac{1}{2}W^{xb}{}_{cd}W^{de}{}_{bf}W^{fc}{}_{gh}W^{gh}{}_{ey}+\\
    &&\quad +W^{xb}{}_{cd}W^{d}{}_{eb}{}^{f}\left(
      W^{ge}{}_{hf}W^{ch}{}_{yg}
      +\frac{1}{2}W^{ce}{}_{gh}W^{gh}{}_{fy}
      -\frac{1}{4}\gd^{[c}_{f}\gd^{e]}_{y}
      W^{ij}{}_{kl}W^{kl}{}_{ij}\right)=\nonumber\\
    &&=-\frac{1}{2}W^{xb}{}_{cd}W^{de}{}_{bf}W^{fc}{}_{gh}W^{gh}{}_{ey}
    +W^{xb}{}_{cd}W^{d}{}_{eb}{}^{f}W^{ge}{}_{hf}W^{ch}{}_{yg}+\\
    &&\quad +\frac{1}{2}W^{xb}{}_{cd}\underbrace{W^{d}{}_{eb}{}^{f}}_{Bianchi}
    W^{ce}{}_{gh}W^{gh}{}_{fy}
    -\frac{1}{8}W^{x}{}_{bcd}W^{bcd}{}_{y}W^{ij}{}_{kl}W^{kl}{}_{ij}=\nonumber\\
    &&=-\frac{1}{2}W^{xb}{}_{cd}W^{de}{}_{bf}W^{fc}{}_{gh}W^{gh}{}_{ey}
    +W^{xb}{}_{cd}W^{d}{}_{eb}{}^{f}W^{eg}{}_{fh}W^{hc}{}_{gy}-\\
    &&\quad -\frac{1}{2}\underbrace{W^{xb}{}_{cd}
      W^{d}{}_{b}{}^{f}{}_{e}}_{lem.\ \ref{t-bianchiComp}}
    W^{ce}{}_{gh}W^{gh}{}_{fy}
    -\frac{1}{2}W^{xb}{}_{cd}W^{df}{}_{eb}W^{ce}{}_{gh}W^{gh}{}_{fy}-\nonumber\\
    &&\quad-\frac{1}{8}W^{x}{}_{bcd}W^{bcd}{}_{y}W^{ij}{}_{kl}W^{kl}{}_{ij}=\nonumber\\
  \end{eqnarray}
  \begin{eqnarray}
    &&=-\underbrace{W^{xb}{}_{cd}W^{de}{}_{bf}W^{fc}{}_{gh}W^{gh}{}_{ey}}_
    {cor.\,\ref{t-irregularChain4}}
    +\underbrace{W^{xb}{}_{cd}W^{d}{}_{eb}{}^{f}W^{eg}{}_{fh}W^{hc}{}_{gy}}_
    {th.\,\ref{t-chain3Id-1}}\\
    &&\quad+\frac{1}{4}\underbrace{W^{x}{}_{cbd}W^{bdf}{}_{e}W^{ce}{}_{gh}
      W^{gh}{}_{fy}}_{lem.\,\ref{t-brokenChain1Id-1}}
    -\frac{1}{8}\underbrace{W^{x}{}_{bcd}W^{bcd}{}_{y}}_
    {th.\,\ref{t-chain3Id-1}}W^{ij}{}_{kl}W^{kl}{}_{ij}=\nonumber\\
    &&=-\frac{1}{4}\gd^{x}_{y}W^{ab}{}_{cd}W^{de}{}_{bf}
    W^{fc}{}_{gh}W^{gh}{}_{ea}
    +\frac{1}{4}\gd^{x}_{y}W^{ab}{}_{cd}W^{d}{}_{eb}{}^{f}
    W^{eg}{}_{fh}W^{hc}{}_{ga}~~~~\label{q-4brokenChainId-2}\\
    &&\quad+\frac{1}{16}\gd^{x}_{y}W^{a}{}_{cbd}W^{bdf}{}_{e}
    W^{ce}{}_{gh}W^{gh}{}_{fa}
    -\frac{1}{32}\gd^{x}_{y}W^{a}{}_{bcd}W^{bcd}{}_{a}
    W^{ij}{}_{kl}W^{kl}{}_{ij}\nonumber
  \end{eqnarray}
  Taking the trace of \refeqn{q-4brokenChainId-2} and multiplying by
  $\frac{1}{4}\gd^{x}_{y}$ gives the lemma.
\end{proof}
It is worth noting that this result may also be found in a slightly
different way. By using only the block- and antisymmetries of a Weyl
candidate, together with Lemma \ref{t-switchId-1} several times, we
have
\begin{eqnarray}\label{q-prodId-2b}
  &&W^{xb}{}_{cd}W^{de}{}_{bf}W^{fgc}{}_{h}W^{h}{}_{egy} =\\
  &&\qquad \frac{1}{4}W^{x}{}_{ebd}W^{bdc}{}_{f}W^{ef}{}_{gh}W^{gh}{}_{cy}
  +W^{xb}{}_{cd}W^{de}{}_{bf}W_{e}{}^{hf}{}_{g}W^{gc}{}_{hy}-\nonumber\\
  &&\qquad -\frac{1}{2}\left(W^{xe}{}_{bd}W^{bd}{}_{cf}W^{chf}{}_{g}W^{g}{}_{ehy}
    +W^{xbc}{}_{d}W^{d}{}_{ebf}W^{ef}{}_{gh}W^{gh}{}_{cy}\right)+\nonumber\\
  &&\qquad +\left(\frac{3}{16}W^{xb}{}_{cd}W^{cd}{}_{yb}-
    \frac{1}{4}W^{x}{}_{bcd}W^{bcd}{}_{y}\right)W^{ij}{}_{kl}W^{kl}{}_{ij}\nonumber
\end{eqnarray}
The lemma now follows by use of Theorem \ref{t-basic4DId-1}, 
\ref{t-chain3Id-1}, \ref{t-brokenChain1Id-1}
together with Corollary \ref{t-distChain1Id-1} on the right-hand-side.
This method requires the Binachi identity only in the final step
(through Corollary \ref{t-distChain1Id-1}). Therefore, it seems likely
that it is possible to generalize Lemma \ref{t-brokenChain2Id-1} to
trace-free block symmetric double $2$-forms.

It is of course possible to break the chain structures in other ways,
and move towards more and more general expressions. It may also be
possible to derive identities for such structures, letting new results
build on previous ones. We shall however not pursue this work here,
since our results are sufficient for our purposes in Section
\ref{s-application}.

\section{Applications}\label{s-application}
\begin{definition}[The Bel-Robinson tensor]
  The \concept{Bel-Robinson tensor} in $n$ dimensions can be defined by
  \begin{eqnarray}\label{q-belRob}
    \se{T}_{abcd} & = &   
    C{}_{aecf}C{}_{b}{}^{e}{}_{d}{}^{f}
    +C{}_{aedf}C{}_{b}{}^{e}{}_{c}{}^{f}
    -\frac{1}{2}g{}_{ab}C{}_{efcg}C{}^{ef}{}_{d}{}^{g}
    -\frac{1}{2}g{}_{cd}C{}_{aefg}C{}_{b}{}^{efg}\nonumber\\
    &&+\frac{1}{8}g{}_{ab}g{}_{cd}C{}_{efgh}C{}^{efgh}
  \end{eqnarray}
\end{definition}
where $C_{abcd}$ denotes the Weyl tensor.  The Bel-Robinson tensor is the
super-energy tensor for the Weyl tensor \cite{Senovilla}. It is of
interest not only in four dimensional general relativity, but has
found applications in, among other places, eleven dimensional
super-gravity theories \cite{Deser}.

It is easy to confirm, using dimensionally dependent identities, that
the Bel-Robinson tensor is trace-free in four dimensions
\cite{Senovilla}, and completely symmetric in, and only in, four and
five dimensions \cite{Edgar02,Senovilla}. Hence, in four dimensions
\begin{eqnarray}
  \se{T}_{abcd} = \se{T}_{(abcd)} \qquad \se{T}^{k}{}_{kcd} = 0
\end{eqnarray}

We are now ready to state one of our main results. This, has been
known for some time in the special case when the metric has a
Lorentzian signature. It was first derived by Debever \cite{Debever}
using principal null directions, and later by Penrose
\cite{Penrose-Rindler-1} using spinor methods. To our knowledge there
does not seem to be a proof for spaces with metrics of arbitrary
signature. However, using dimensionally dependent identities, we are
able to go beyond this special case and give a proof for metrics
\emph{of any signature} in four dimensions.

\begin{theorem}\label{t-quadraticBR-1}
  In four dimensions
  \begin{equation}\label{q-quadraticBRId-1}
    \se{T}_{abcy}\se{T}^{abcx} = \frac{1}{4}\gd_{y}^{x}\se{T}_{abcd}\se{T}^{abcd}
  \end{equation}
\end{theorem}

\begin{proof}
  Simple substitution yields,
  \begin{eqnarray}\label{q-BR-LHS}
    &&\se{T}_{abcy}\se{T}^{abcx} -
    \frac{1}{4}\gd_{y}^{x}\se{T}_{abcd}\se{T}^{abcd} = \nonumber\\
    &&\quad 2C^{xb}{}_{cd}C^{de}{}_{bf}C^{fg}{}_{eh}C^{hc}{}_{gy}
    +2C^{xb}{}_{cd}C^{de}{}_{bf}C^{fgc}{}_{h}C^{h}{}_{egy}-\nonumber\\
    &&\quad -2C^{ab}{}_{cd}C^{cd}{}_{eb}C^{e}{}_{gh}{}^{x}C_{a}{}^{gh}{}_{y}
    -C^{xb}{}_{cd}C^{cd}{}_{eb}C^{ef}{}_{gh}C^{gh}{}_{yf}+\nonumber\\
    &&\quad +\frac{1}{2}C_{abcd}C^{abcd}C^{ex}{}_{gh}C^{gh}{}_{ey}
    +\frac{1}{4}\gd_{y}^{x}C^{ab}{}_{cd}C^{cd}{}_{eb}C^{ef}{}_{gh}C^{gh}{}_{af}-\nonumber\\
    &&\quad -\frac{1}{16}\gd_{y}^{x}C_{abcd}C^{abcd}C_{efgh}C^{efgh}-\nonumber\\
    &&\quad -\frac{1}{4}\gd_{y}^{x}
    \Big(2C^{ab}{}_{cd}C^{de}{}_{bf}C^{fg}{}_{eh}C^{hc}{}_{ga}
    +2C^{ab}{}_{cd}C^{de}{}_{bf}C^{fgc}{}_{h}C^{h}{}_{ega}-\nonumber\\
    &&\qquad \qquad -2C^{ab}{}_{cd}C^{cd}{}_{eb}C^{ef}{}_{gh}C^{gh}{}_{af}
      +\frac{1}{4}C_{abcd}C^{abcd}C_{efgh}C^{efgh}\Big).
  \end{eqnarray}
  Using Theorem \ref{t-basic4DId-1} several times on the
  right-hand-side, this simplifies to
  \begin{eqnarray}\label{q-BR-LHS-4D}
    && \se{T}_{abcy}\se{T}^{abcx} -
    \frac{1}{4}\gd_{y}^{x}\se{T}_{abcd}\se{T}^{abcd} = \nonumber\\
    &&\quad 2C^{xb}{}_{cd}C^{de}{}_{bf}C^{fg}{}_{eh}C^{hc}{}_{gy}
    +2C^{xb}{}_{cd}C^{de}{}_{bf}C^{fgc}{}_{h}C^{h}{}_{egy}-\nonumber\\
    &&\quad -\frac{1}{4}\gd_{y}^{x}\left(
      2C^{ab}{}_{cd}C^{de}{}_{bf}C^{fg}{}_{eh}C^{hc}{}_{ga}
      +2C^{ab}{}_{cd}C^{de}{}_{bf}C^{fgc}{}_{h}C^{h}{}_{ega}\right).
  \end{eqnarray}
  The first term and the third term cancel by Theorem
  \ref{t-chain2Id-1}, and the second and the fourth term cancel
  by Lemma \ref{t-brokenChain2Id-1}, and the theorem follows.
\end{proof}
Since we have used only the algebraic properties, it is obvious that
the theorem is true not just for the Weyl tensor, but any Weyl candidate.

We have noted above that $\se{T}_{abcd}$ is completely symmetric in
five as well as in four dimensions. So, a natural question to arise is
whether a similar identity exists in a five dimensional space or not?
One can show however, that this question is answered in the negative
\cite{Edgar03}.  

In the next application we turn our attention to the super-energy
tensor of the Riemann tensor, also known as the Bel tensor. It may be
defined in complete analogy with equation \refeqn{q-belRob} by
exchanging $C_{abcd}$ for $R_{abcd}$ \cite{Senovilla}. The Bel tensor
may be decomposed into four parts,
\begin{equation}\label{q-belDecomp}
  \se{B}_{abcd} = \se{T}_{abcd} + \se{Q}_{abcd} + \se{G}_{abcd} +
  \se{E}_{abcd}
\end{equation}
where $\se{T}_{abcd}$ is the Bel-Robinson tensor. We will address
$\se{Q}_{abcd}$ and $\se{G}_{abcd}$ shortly. $\se{E}_{abcd}$ is a
fairly complicated expression including products of the metric and the
trace-free Ricci tensor. We will not discuss this tensor further, but
we note that in four dimensions, it satisfies \cite{Edgar03} the
relation
\begin{equation}\label{q-quadraticSuperE}
  \se{E}_{abcy}\se{E}^{abcx} =
  \frac{1}{4}\gd^{x}_{y}\se{E}_{abcd}\se{E}^{abcd}
\end{equation}
In four dimensions \cite{Bonilla-Senovilla,Edgar03},
\begin{eqnarray}
  &&\se{Q}_{abcd} = \frac{R}{6}\left(C_{acbd}+C_{adbc}\right)\\
  &&\se{G}_{abcd} = \frac{R^2}{144}\left(
    4g{}_{a(c}g{}_{d)b}-g{}_{ab}g{}_{cd}
  \right)
\end{eqnarray}
where $R$ denotes the Ricci scalar. They are both easily shown
\cite{Edgar03} to satisfy a relation analogous to
\refeqn{q-quadraticSuperE}.

From a physical point of view it is reasonable to group the last two
terms of equation \refeqn{q-belDecomp} together, as done by Bonilla
and Senovilla \cite{Bonilla-Senovilla},
\begin{equation}\label{q-tensorDecomp}
  \se{B}_{abcd} = \se{T}_{abcd} + \se{Q}_{abcd} + \se{M}_{abcd}
\end{equation}

The structure of the spinor formulation of the Bel tensor
\cite{Bergqvist99} however, suggests another way of grouping the terms
of equation \refeqn{q-belDecomp},
\begin{equation}\label{q-spinorDecomp}
  \se{B}_{abcd} = \se{X}_{abcd} + \se{E}_{abcd}
\end{equation}
and hence,
\begin{equation}
  \se{X}_{abcd} = \se{T}_{abcd} + \se{Q}_{abcd} + \se{G}_{abcd}.
\end{equation}
Since $\se{E}_{abcd}$ satisfies \refeqn{q-quadraticSuperE}, it is
natural to ask whether there is a similar relation for $\se{X}_{abcd}$
or not. As our next theorem shows, this is indeed the case.
\begin{theorem}\label{t-quadraticSuperE}
  In four dimensions
  \begin{equation}
    \se{X}_{abcy}\se{X}^{abcx} =
    \frac{1}{4}\gd^{x}_{y}\se{X}_{abcd}\se{X}^{abcd}
  \end{equation}
\end{theorem}
\begin{proof}
  \begin{eqnarray}\label{q-quadraticSuperX}
    &&\se{X}_{abcy}\se{X}^{abcx} =
    \left(\se{T}_{abcy}+\se{Q}_{abcy}+\se{G}_{abcy}\right)
    \left(\se{T}^{abcx}+\se{Q}^{abcx}+\se{G}^{abcx}\right)=\nonumber\\
    &&\quad=
    \se{T}_{abcy}\se{T}^{abcx}+\se{Q}_{abcy}\se{Q}^{abcx}+
    \se{G}_{abcy}\se{G}^{abcx}+\nonumber\\
    &&\qquad+\se{T}_{abcy}\se{Q}^{abcx}+\se{Q}_{abcy}\se{T}^{abcx}
  \end{eqnarray}
  The cross-terms with $\se{G}_{abcd}$ vanishes since $\se{G}_{abcd}$
  is essentially the metric, and $\se{T}_{abcd}$ and $\se{Q}_{abcd}$
  are trace-free. For the last term, we have
  \begin{eqnarray}
    &&\se{Q}_{abcy}\se{T}^{abcx} =\\
    &&\quad= \frac{1}{3}R\left(
      C{}^{xb}{}_{cd}C{}^{de}{}_{bf}C{}_{e}{}^{fc}{}_{y}
      -C{}^{xb}{}_{cd}C{}^{de}{}_{bf}C{}^{fc}{}_{ey}
      +\frac{1}{2}C{}^{xb}{}_{cy}C{}_{bdef}C{}^{cdef}
    \right).\nonumber
  \end{eqnarray}
  The first term is a twisted chain of the zeroth kind, and hence we
  may use Corollary \ref{t-distChain1Id-1}. The second term is a chain of
  the first kind and Theorem \ref{t-chain2Id-1} may be applied. The
  third term vanishes by Theorem \ref{t-basic4DId-1} and the
  trace-free property of $C_{abcd}$. Hence,
  \begin{equation}\label{q-crossTQ-1}
    \se{Q}_{abcy}\se{T}^{abcx} =
    \frac{1}{12}\gd^{x}_{y}R\left(
      -\frac{1}{2^2}\chain{0}{C}{3}{ac}{ac}
      -\chain{1}{C}{3}{ab}{ab}
    \right) =
    \frac{1}{4}\gd^{x}_{y}\se{Q}_{abcd}\se{T}^{abcd}.
  \end{equation}
  A similar argumet holds for the penultimate term of
  \refeqn{q-quadraticSuperX}. Therefore, by virtue of Theorem
  \ref{t-quadraticBR-1}, and analogous properites of
  $\se{Q}_{abcd}$ and $\se{G}_{abcd}$, we have
  \begin{eqnarray}
    &&\se{X}_{abcy}\se{X}^{abcx} =\nonumber\\
    &&\quad=\frac{1}{4}\gd^{x}_{y}\left(
      \se{T}_{abcd}\se{T}^{abcd}+\se{Q}_{abcd}\se{Q}^{abcd}+
      \se{G}_{abcd}\se{G}^{abcd}+2\se{Q}_{abcd}\se{T}^{abcd}
    \right)\nonumber\\
    &&\quad=\frac{1}{4}\gd^{x}_{y}\se{X}_{abcd}\se{X}^{abcd}
  \end{eqnarray}
  as desired.
\end{proof}
It is also possible to derive a quadratic identity for the full Bel
tensor. It has a slightly different structure however, and is in a
sense trivial \cite{Edgar03}. 

\section{Summary and discussion}
Our main results, besides Theorem \ref{t-quadraticBR-1} and Theorem
\ref{t-quadraticSuperE}, have been the chain identities
\begin{eqnarray}
  \chain{0}{U}{m}{xb}{yb} &=& \frac{1}{4}\gd^{x}_{y}\chain{0}{U}{m}{ab}{ab}\\
  \chain{1}{U}{m}{xb}{by} &=& \frac{1}{4}\gd^{x}_{y}\chain{1}{U}{m}{ab}{ba}\\
  \chain{t1}{V}{m}{xb}{by} &=& \frac{1}{4}\gd^{x}_{y}\chain{t1}{V}{m}{ab}{ba}\\
  \chain{nt0}{W}{m}{wx}{yz} &=& \frac{1}{2^n}\chain{0}{W}{m}{wx}{yz}
\end{eqnarray}
where $U$ is a trace-free (2,2)-form, $V$ in addition enjoys block
symmetry, and $W$ is a Weyl candidate. We have also seen that the
chain structures may be broken in different ways, and that it is
possible to find identities for the resulting structures.

There are several possible directions to pursue with those chain-like
structures. One way is to stay with four index tensors in a four
dimensional space, and investigate more and more broken structures.
This sort of work may be useful in the study of relations between
scalar invariants of, e.g., the Weyl tensor
\cite{Harvey95,Jack86}.

Another direction is to start working with tensors that are \emph{not}
trace-free, and examine what the identities look like for those. Such
calculations are expected to be lengthy and proper computer tools will
be invaluable. 

Yet another direction is to increase the number of dimensions and the
number of indices of the participating tensors. This prospect looks
promising in that Lovelock proved \cite{Lovelock70} a set of
identities of the form,
\begin{eqnarray}
  &&S^{a_1\ldots\ a_{m-1}x}{}_{b_1\ldots\ b_m}
  S^{b_1\ldots\ b_m}{}_{a_1\ldots\ a_{m-1}y} =\nonumber\\
  &&\qquad \frac{1}{n}\gd_{y}^{x}
    S^{a_1\ldots\ a_m}{}_{b_1\ldots\ b_m}
  S^{b_1\ldots\ b_m}{}_{a_1\ldots\ a_m}
\end{eqnarray}
where $S$ is a trace-free (m,m)-form in an $n=2m$ dimensional space.

It therefore seems very likely that there exist chain identities in
higher dimensions for tensors with more indices in direct analogy with
the relations found in Section \ref{s-chains}. However, it will be
possible to form more than just two kinds of chains.

Suppose, for instance, that $n=6$. Then we are able to form three
different kinds of chains,
\begin{eqnarray}
  &&S^{xb_1c_1}{}_{d_1e_1f_1}S^{f_1e_1d_1}{}_{d_2e_2f_2}
  S^{f_2e_2d_2}{}_{d_3e_3f_3}\ldots
  S^{f_{k-1}e_{k-1}d_{k-1}}{}_{c_1b_1y}\\
  &&S^{xb_1c_1}{}_{d_1e_1f_1}S^{f_1e_1c_2}{}_{c_1e_2f_2}
  S^{f_2e_2c_3}{}_{c_2e_3f_3}\ldots
  S^{e_{k-1}f_{k-1}d_1}{}_{c_{k-1}b_1y}\\
  &&S^{xb_1c_1}{}_{d_1e_1f_1}S^{f_1b_2c_2}{}_{c_1b_1f_2}
  S^{f_2b_3c_3}{}_{c_2b_2f_3}\ldots
  S^{f_{k-1}e_1d_1}{}_{b_{k-1}c_{k-1}y}
\end{eqnarray}
It may even be possible to find analogies to the twisted or more
broken chain structures of Sections \ref{s-blockSymChain} and
\ref{s-weylChains}; in particular if we consider tensors with extra
symmetries, e.g., block symmetry or the Bianchi like property
$S_{ab[cdef]} = 0$. However, such possibilities needs further
investigation.

\section*{Acknowledgements}
I wish to give special thanks to Prof. Brian Edgar for introducing me
to dimensionally dependent tensor identities, and his proposal to
investigate the quadratic identity for the Bel-Robinson tensor.
  
I also wish thank Tekn D. Anders Höglund for letting me use a pre-release
of his tensor manipulation program ``Tensign'' \cite{Tensign}. Without
it, some of the calculations presented here would have been next to
impossible.

\appendix

\end{document}